\pdfoutput=1
\documentclass[submission, PhysProc]{SciPost}

\hypersetup{
    colorlinks,
    linkcolor={red!50!black},
    citecolor={blue!50!black},
    urlcolor={blue!80!black}
}

\usepackage[bitstream-charter]{mathdesign}
\DeclareSymbolFont{usualmathcal}{OMS}{cmsy}{m}{n}
\DeclareSymbolFontAlphabet{\mathcal}{usualmathcal}

\begin{document}

\begin{center}{\Large \textbf{
Study of the Hyperon-Nucleon Interaction via Final-State Interactions in Exclusive Reactions. \\
}}\end{center}

\begin{center}
Nicholas Zachariou\textsuperscript{1$\star$}, Daniel Watts\textsuperscript{1}, and Yordanka Ilieva\textsuperscript{2}\\
for the CLAS collaboration.
\end{center}

\begin{center}
{\bf 1} University of York
{\bf 2} University of South Carolina
\\
${}^\star$ {\small \sf nick.zachariou@york.ac.uk}
\end{center}

\begin{center}
\today
\end{center}


\definecolor{palegray}{gray}{0.95}
\begin{center}
\colorbox{palegray}{
  \begin{tabular}{rr}
  \begin{minipage}{0.05\textwidth}
    \includegraphics[width=14mm]{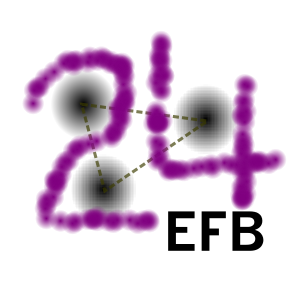}
  \end{minipage}
  &
  \begin{minipage}{0.82\textwidth}
    \begin{center}
    {\it Proceedings for the 24th edition of European Few Body Conference,}\\
    {\it Surrey, UK, 2-4 September 2019} \\
    \end{center}
  \end{minipage}
\end{tabular}
}
\end{center}

\section*{Abstract}
{\bf
A novel approach that allows access to long-sought information on the Hyperon-Nucleon (YN) interaction was developed by producing a hyperon beam  within a few-body nuclear system, and studying final-state interactions. The determination of polarisation observables, and specifically the beam spin asymmetry,  in exclusive reactions allows a detailed study of the various final-state interactions and provides us with the tools needed to isolate kinematic regimes where the YN interaction dominates. High-statistics data collected using the CLAS detector housed in Hall-B of the Thomas Jefferson laboratory allows us to obtain a large set of polarisation observables and place stringent constraints on the underlying dynamics of the YN interaction.  }


\section{Introduction}
\label{sec:intro}
One of the main goal of nuclear physics is to obtain a comprehensive picture of the strong interaction, which can be accessed by introducing the strangeness degree of freedom in the, now well-understood, nucleon-nucleon ($NN$) interaction. The $NN$ interaction has a long history of detailed studies, and currently phenomenological approaches can describe observed phenomena with high accuracy. On the other hand, the interaction between Hyperons (hadrons with one or more strange quarks) and Nucleons ($YN$) is very poorly constrained, mainly due to difficulties associated with performing high-precision scattering experiments involving short-lived hyperon beams. 
Because of these difficulties, complimentary approaches, including studies of hypernuclei and final-state interaction (FSI), have been developed to provide  indirect access to information on the hyperon-nucleon interaction.  Final state interactions in exclusive hyperon photoproduction reactions off deuterium impart an excellent tool to study the bare $YN$ interaction in an approach that is free from medium modifications and many-body effects. 

\begin{figure}[h]
\centering
\includegraphics[width=0.7\linewidth]{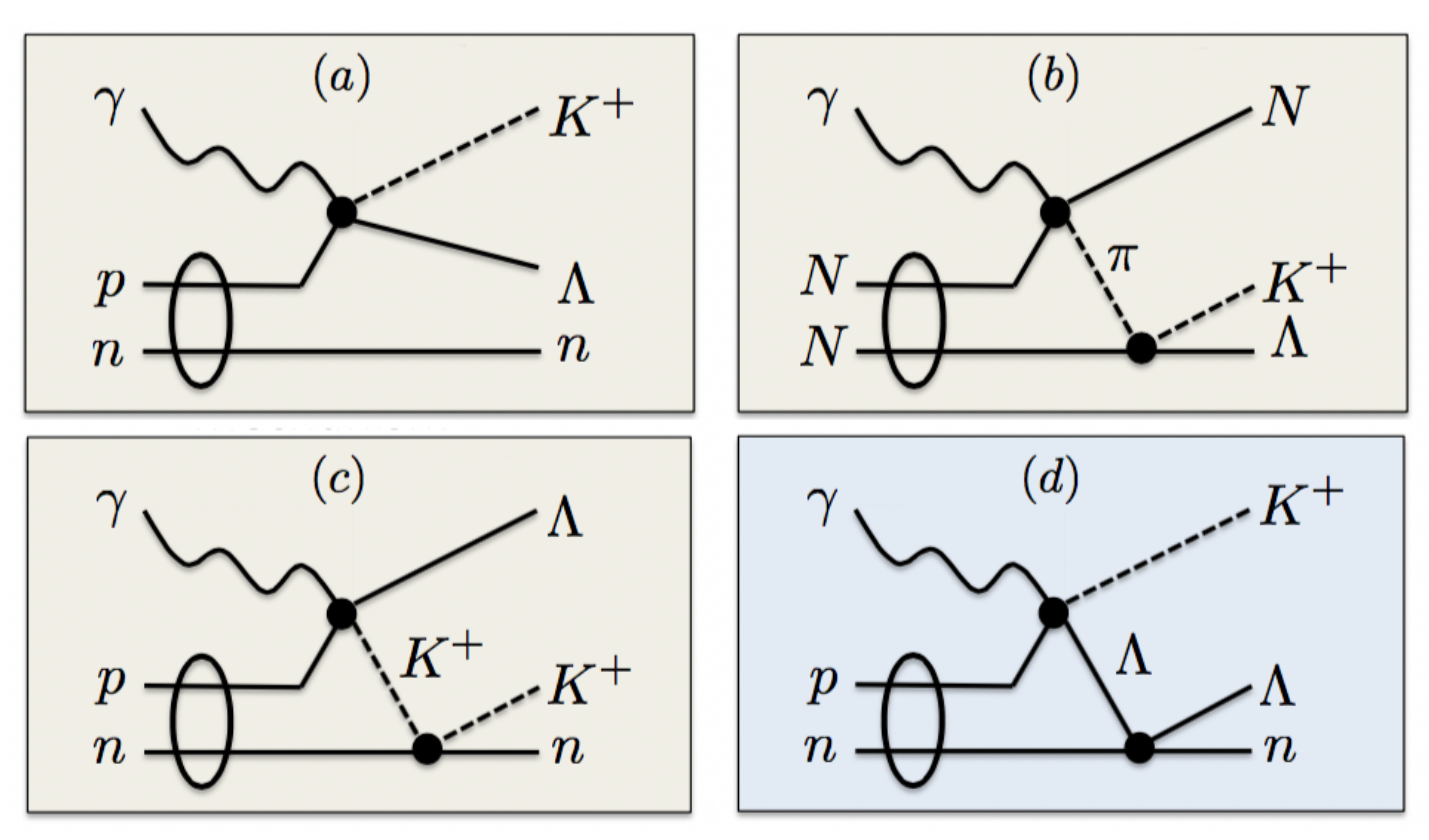}%
\caption{\label{Fig:Reactions} Four main mechanisms that contribute to the reaction $\gamma d \rightarrow K^+\Lambda n$ according to theoretical models~\cite{Ref:Salam,Ref:Laget}: (a) quasi-free $\Lambda K^+$ photoproduction on the proton; (b) pion mediated production; (c) $K^+$ rescattering off the spectator neutron; (d) $\Lambda$ rescattering off the  spectator neutron.}
\end{figure}

Available models~\cite{Ref:Salam,Ref:Laget} indicate that four mechanisms contribute in the exclusive reaction $\gamma d\to K^+\Lambda n$ (see Fig.~\ref{Fig:Reactions}): (1) the quasi-free mechanism, which dominates the reaction cross section, (2) the pion mediated mechanism, (3) the kaon rescattering, and (4) the hyperon rescattering mechanism.
 The exclusivity of the reaction allows us to place kinematical constraints that minimise contributions from the quasi-free reaction, enabling detailed investigation of FSI. Linearly and circularly polarised photon beams in combination with the self-analyticity of hyperons give access to a large set of polarisation observables that are crucial for constraining the dynamics of the $YN$ interaction. This is illustrated by the most comprehensive model for this reaction, which uses two $YN$ potentials (Nijmegen NSC89 and NSC97f -- both of which correctly predict the hypertrition binding energy)~\cite{Ref:Miyakawa} and provides calculations of the polarised differential cross section, allowing predictions for a set of polarisation observables. These calculations predict a strong sensitivity between the two main potentials, indicating that polarisation observables are crucial in our current understanding of the $YN$ interaction. A determination of a large set of polarisation observables will allow a model-dependent interpretation of the data, in which various FSI contribute. The beam-spin asymmetry, $\Sigma$, is a critical observable that allows direct insight on contributions from the different FSI.  

\section{Experimental setup}
\begin{figure}[h]
\includegraphics[width=\linewidth]{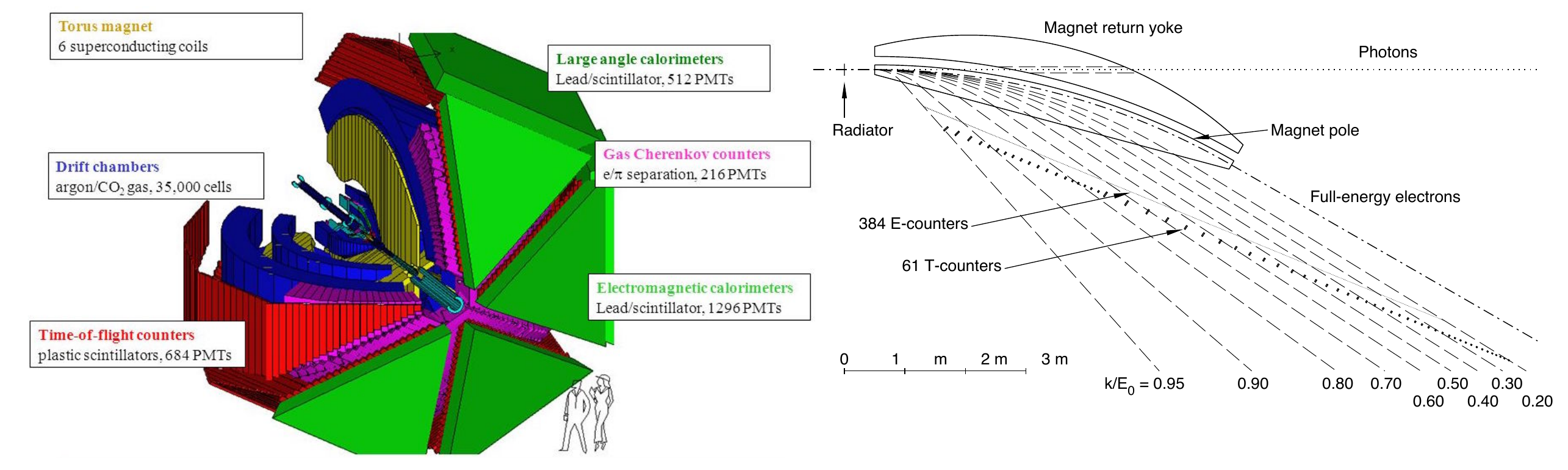}%
\caption{\label{Fig:CLAS} Left: A schematic of the CLAS detector. The kidney-shaped superconducting coils are shown in yellow, drift
chambers in blue, \u{C}erenkov counters in magenta, time-of-flight scintillators in red, and electromagnetic calorimeters in green. Right: The Hall-B photon tagging system located upstream of the CLAS detector.}
\end{figure}

Recent developments in accelerator and detector technologies, allowed the collection of a  large data sample of the exclusive reaction $\vec{\gamma} d\to K^+\Lambda n$ utilising the CEBAF~\cite{Ref:CEBAF} Large Acceptance Spectrometer (CLAS)~\cite{Ref:CLAS} and the tagger spectrometer housed in Hall-B of the Thomas Jefferson Laboratory (JLab). The CLAS detector, which is comprised of a time-of-flight detector system, drift chambers, a superconducting magnet that produced a toroidal magnetic field, and electromagnetic calorimeters,    provided us with an efficient detection of charged particles over a large fraction of the full solid angle. A schematic of the CLAS detector is shown in the left panel of Fig.~\ref{Fig:CLAS}.

\begin{figure}[h]
\centering
\includegraphics[width=0.87\linewidth]{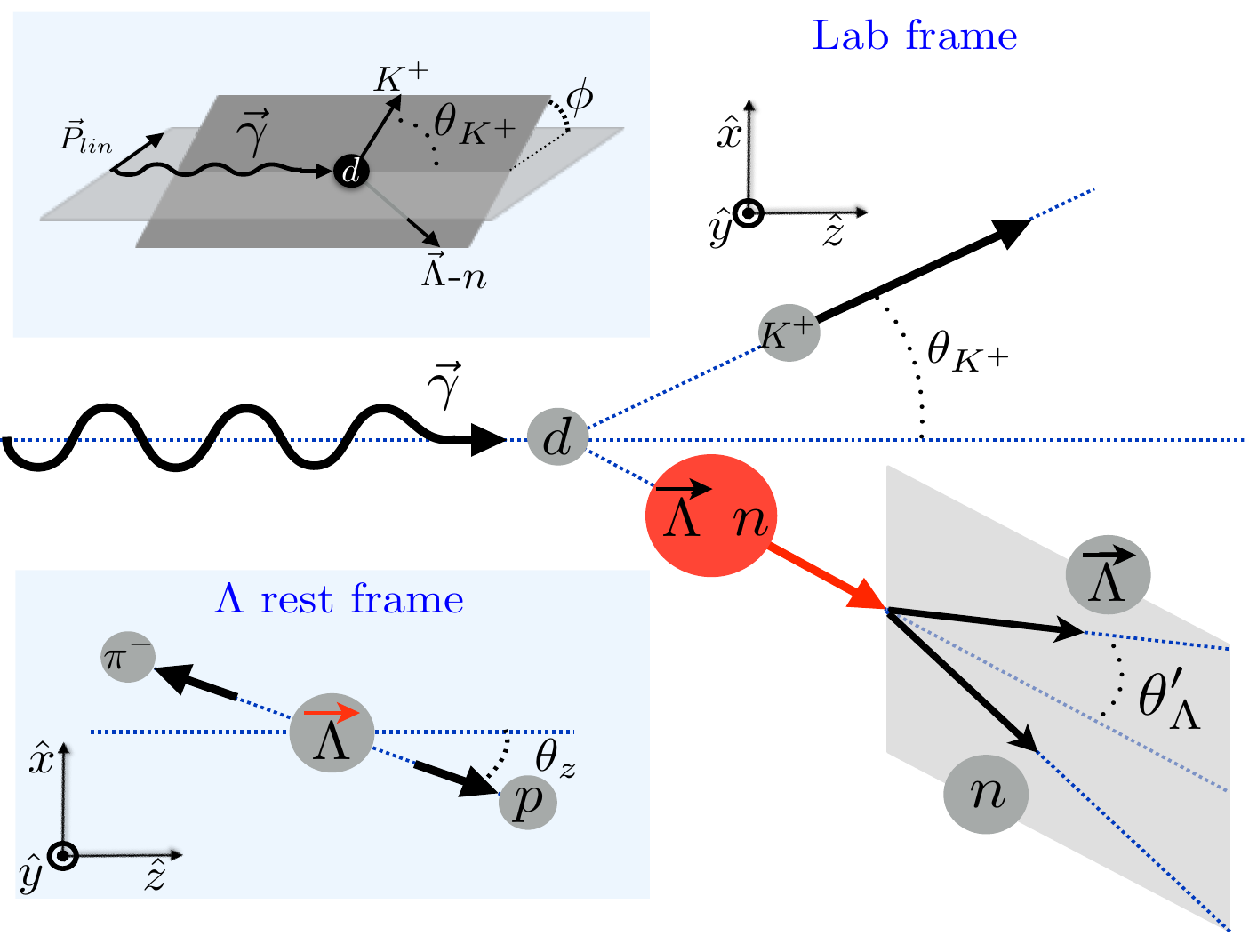}\hspace{0.5pc}%
\caption{\label{Fig:ReacPlane} Reaction plane definition for $\gamma d\rightarrow K^+\Lambda n$. The upper left panel shows the definition of the kaon azimuthal angle $\phi$ with which $\Sigma$ produces a modulation (see Eq.~(\ref{Eq1})). The lower left panel shows the definition of $\theta_z$ (and correspondingly $\theta_x$) with which the double-polarization observables produce modulations. The main figure defines the kaon laboratory polar angle $\theta_{K^+}$, and the hyperon angle $\theta_{\Lambda'}$, which is with respect to an axis pointing along the momentum of the YN pair. }
\end{figure}

Data for this analysis were collected from one of the largest photoproduction experiments conducted during the CLAS6 era (experiment E06-103~\cite{Ref:g13}), utilising both linearly and a circularly polarised tagged photon beams incident on a 40~cm long deuterium target. The tagger spectrometer (right panel of Fig.~\ref{Fig:CLAS}) allowed the tagging of photons with energies between 20 and 95\% of the incident electron energy. A linearly polarised photon beam between 0.7 and 2.3 GeV produced via the coherent bremsstrahlung technique (electrons incident on a diamond radiator), gave access to the beam spin asymmetry $\Sigma$, as well as the polarisation transfer observables $O_x$ and $O_z$. On the other hand, photons produced using an amorphous radiator and a polarised electron beam, were circularly polarised and gave access to the double polarisation observables $C_x$ and $C_z$. Equation~(\ref{Eq1}) shows how the cross section of the reaction $\gamma d\rightarrow K^+\Lambda n$ depends on the set of polarisation observables this work focuses on:
\begin{eqnarray}
\frac{d\sigma}{d\Omega}&=&\left(\frac{d\sigma}{d\Omega} \right)_0[1-P_{lin}\Sigma\cos2\phi+\alpha\cos\theta_x(-P_{lin}O_x\sin2\phi-P_{circ}C_x)     \nonumber \\
&&-\alpha\cos\theta_y(-P_y+P_{lin}T\cos2\phi)-\alpha\cos\theta_z(P_{lin}O_z\sin2\phi+P_{circ}C_z)],\label{Eq1}
\end{eqnarray}
where $\phi$ is the azimuthal angle of the kaon, and $P_{lin}$ and $P_{circ}$ is the degree of linear and circular polarisation respectively, and $\alpha$ is the self-analsing power of the $\Lambda$ hyperon equal to 0.75~\cite{Ref:PDG}.  Figure~\ref{Fig:ReacPlane} shows the frame definition used to determine all relevant polarisation observables. 

\section{Analysis}
\begin{figure}[h]
\centering
\includegraphics[width=0.97\linewidth]{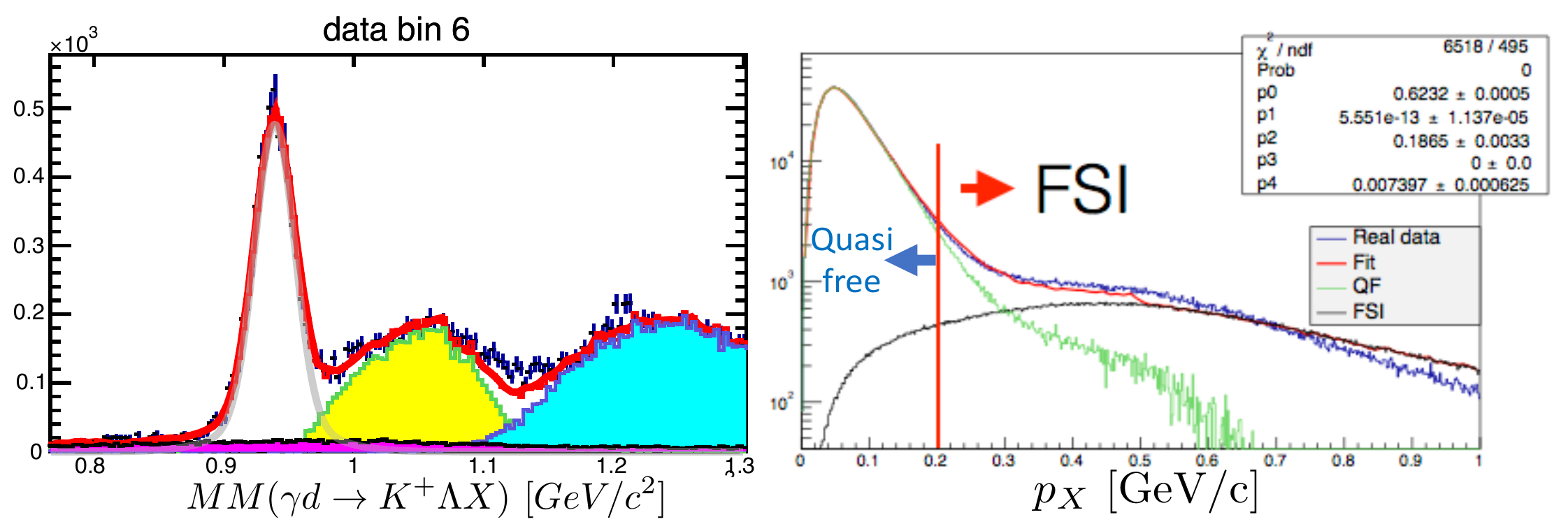}\hspace{0.5pc}%
\caption{\label{Fig:MMPX} Left: Missing-mass distribution of $\gamma d \rightarrow K^+\Lambda X$ indicated the various sources of background ($\Sigma$ channel with yellow, $\Sigma^*$ channel with cyan, and accidentals with magenta). Right: Missing momentum of the reaction indicating the typical cut applied to select quasi-free or FSI- dominated samples. }
\end{figure}

The reaction $\gamma d\rightarrow K^+\Lambda n$ was fully reconstructed with the detection of all charged-tracks in the final state, utilising the large branching ratio of $\Lambda\to p\pi^-$ (63.9\%). Particle identification was done based on time-of-flight and drift chamber information, and misidentified kaons were discarded with a requirement on the proton-pion invariant mass. Finally, the missing neutron was identified utilising the tagged photon-beam by constructing the missing-mass, $M_X$, of the reaction $\gamma d\rightarrow K^+\Lambda X$. Background contributions, mainly from $\Sigma^0$ (which decays into a $\Lambda$ and a $\gamma$), were accounted for  using a comprehensive event generator and a realistic detector simulation based on the GEANT package~\cite{Ref:GSIM}. The left panel of Fig.~\ref{Fig:MMPX} shows an example of the missing-mass distribution for a specific kinematic bin, indicating the various background contributions. Quasi-free and FSI-dominated samples were determined by selecting events with neutron momenta (reconstructed from the missing-momentum of the reaction $\gamma d\rightarrow K^+\Lambda X$) below and above 200~MeV/c, as indicated in the right panel of Fig.~\ref{Fig:MMPX}. 
The polarisation observables were determined using the maximum likelihood technique. Specifically, the likelihood function was established from the cross section of the reaction (see Eq.~(\ref{Eq1})). This allowed for a simultaneous extraction of a set of polarisation observables by maximising the log-likelihood calculated using the linearly and circularly polarised data. 

As indicated before, the beam-spin asymmetry plays a crucial role in this study, since it provides us with vital information that allows us to identify kinematic regimes where specific FSI mechanisms dominate. This observable can be determined using the azimuthal distribution of kaons, $\Sigma_{K^+}$, hyperons, $\Sigma_{\Lambda}$, and neutrons, $\Sigma_{n}$. In a quasi-free dominated sample, where the neutron is a spectator, the hyperon and kaon azimuthal distribution are expected to result in the same beam-spin asymmetries (i.e. $\Sigma_{K^+}=\Sigma_{\Lambda}$), which would also be consistent to the beam-spin asymmetry determined using a free-proton target (any contributions from initial-state effects are expected to be small). Moreover, the neutron beam-spin asymmetry $\Sigma_{n}$ is expected to be consistent with zero as it does not participate in the reaction. In the case of a sample that is dominated by the pion-mediate reaction, the beam-spin asymmetries $\Sigma_{K^+}$ and $\Sigma_{\Lambda}$ are expected to be consistent with zero, where the neutron beam-spin asymmetry, $\Sigma_{n}$, is expected to be consistent with the well-known pion beam-spin asymmetries of the reactions $\gamma n\to\pi^0 n$ or $\gamma p\to\pi^+ n$. For a kaon-rescattering dominated  sample, the neutron beam-spin asymmetry, $\Sigma_{n}$, is expected to be consistent with zero, where the hyperon beam-spin asymmetry, $\Sigma_{\Lambda}$ is expected to be consistent with the quasi-free value. In this case, the kaon beam-spin asymmetry $\Sigma_{K^+}$, should be diluted with respect to the quasi-free value due to the secondary scattering process. A similar situation is expected in a hyperon-rescattering dominated sample, where the beam-spin asymmetry determined using the hyperon azimuthal distribution is expected to be diluted with respect to the quasi-free value, where the $\Sigma_{K^+}$ should be consistent with the quasi-free value and $\Sigma_{n}$ consistent to zero. This information was extensively studied using generated samples from our comprehensive event generator that includes rescattering processes in a simplified two-step approach.

\section{Results and discussion}
Generated samples allowed us to study in great detail the various contributing FSI in a controlled manner.
 \begin{figure}[h!]
\centering
\includegraphics[width=0.47\linewidth]{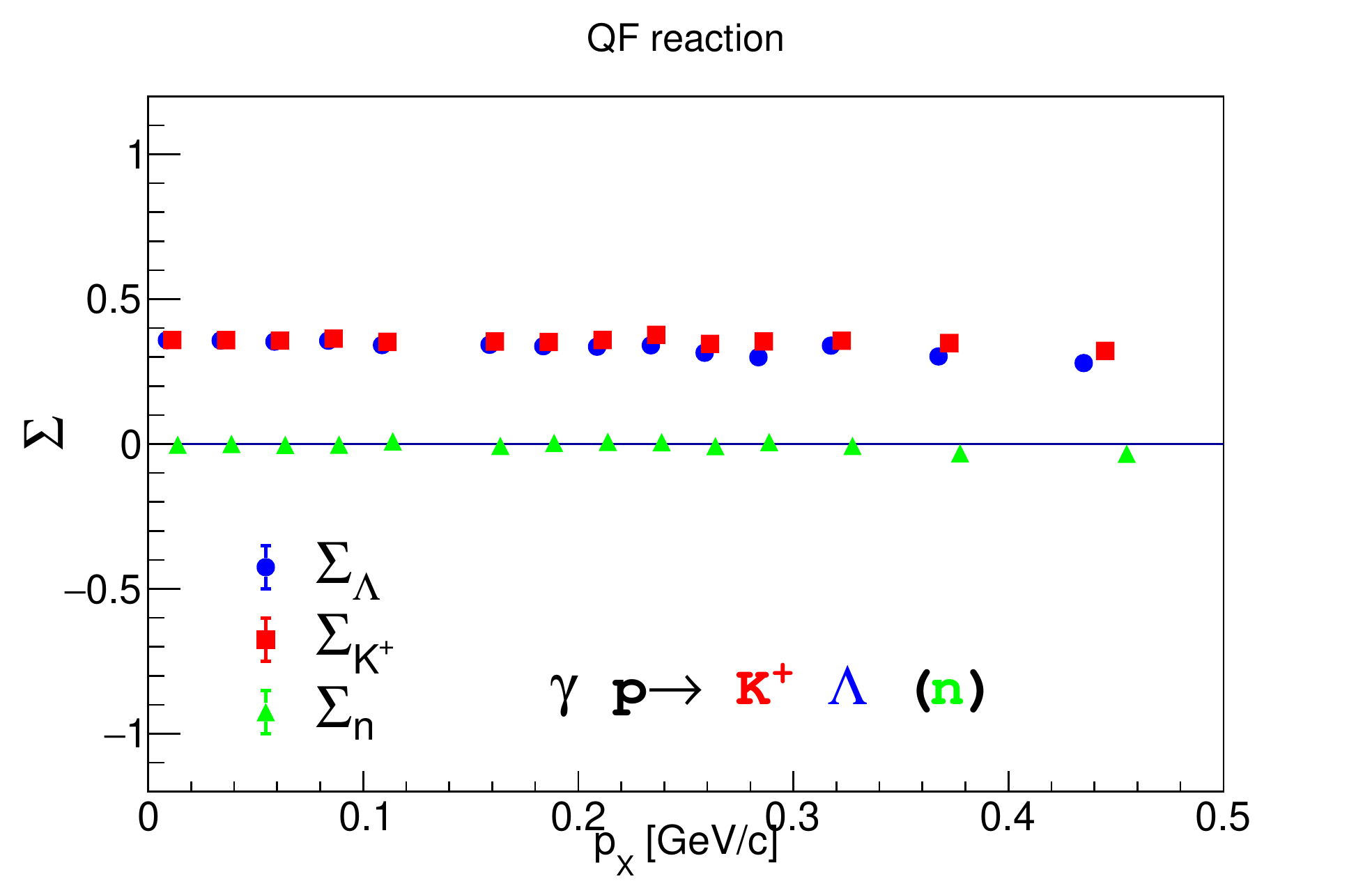}
\includegraphics[width=0.47\linewidth]{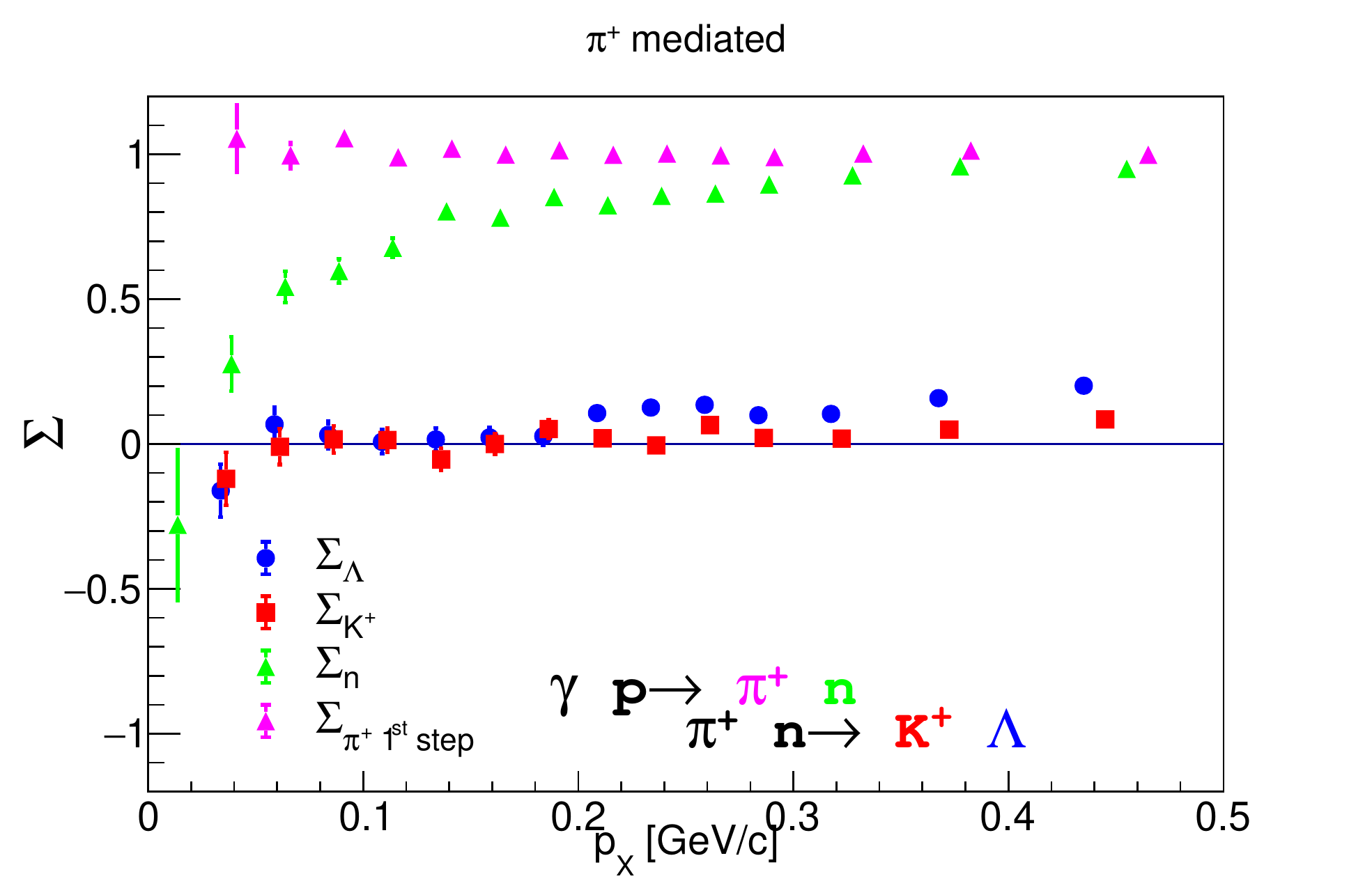}
\includegraphics[width=0.47\linewidth]{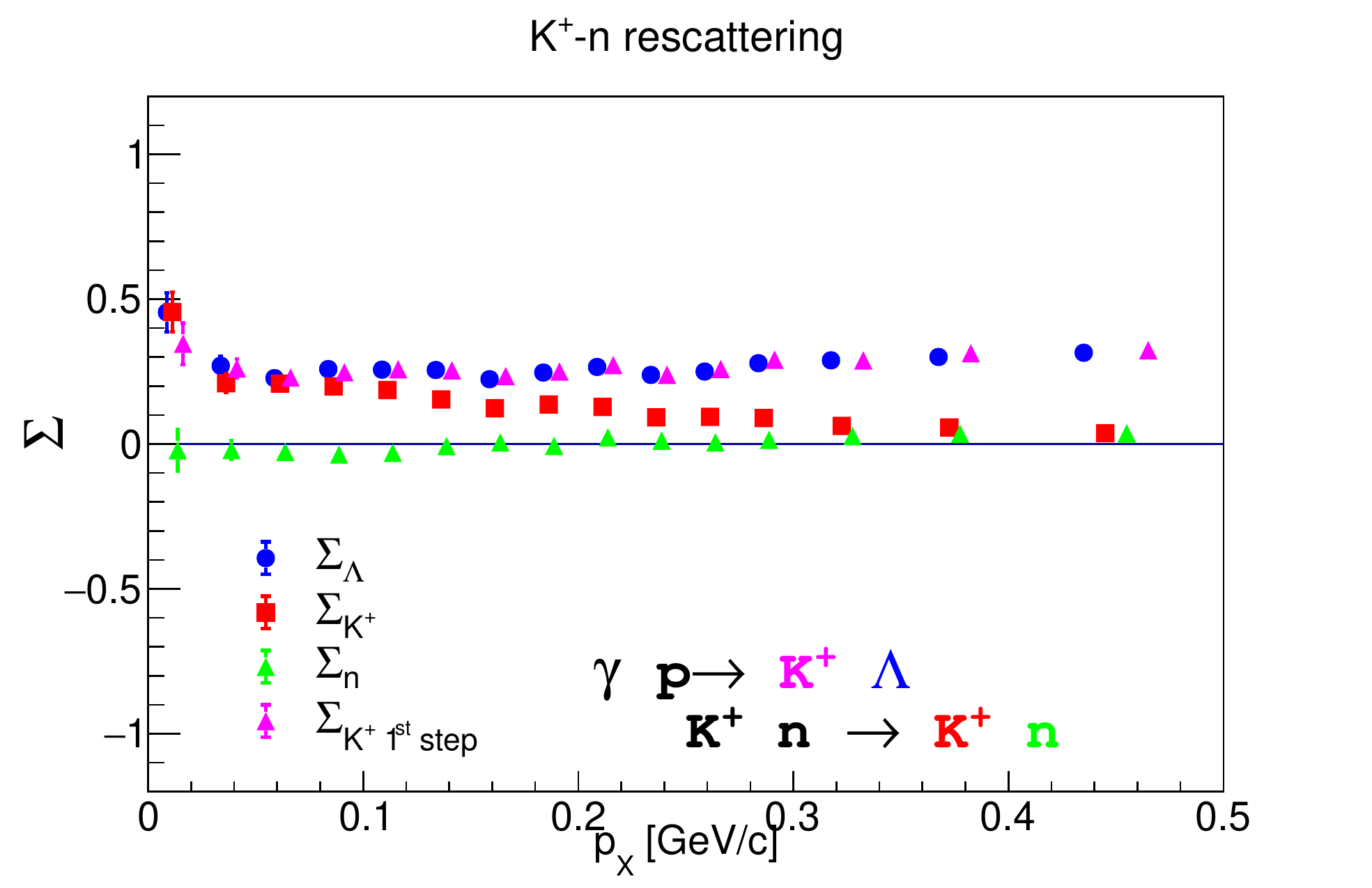}
\includegraphics[width=0.47\linewidth]{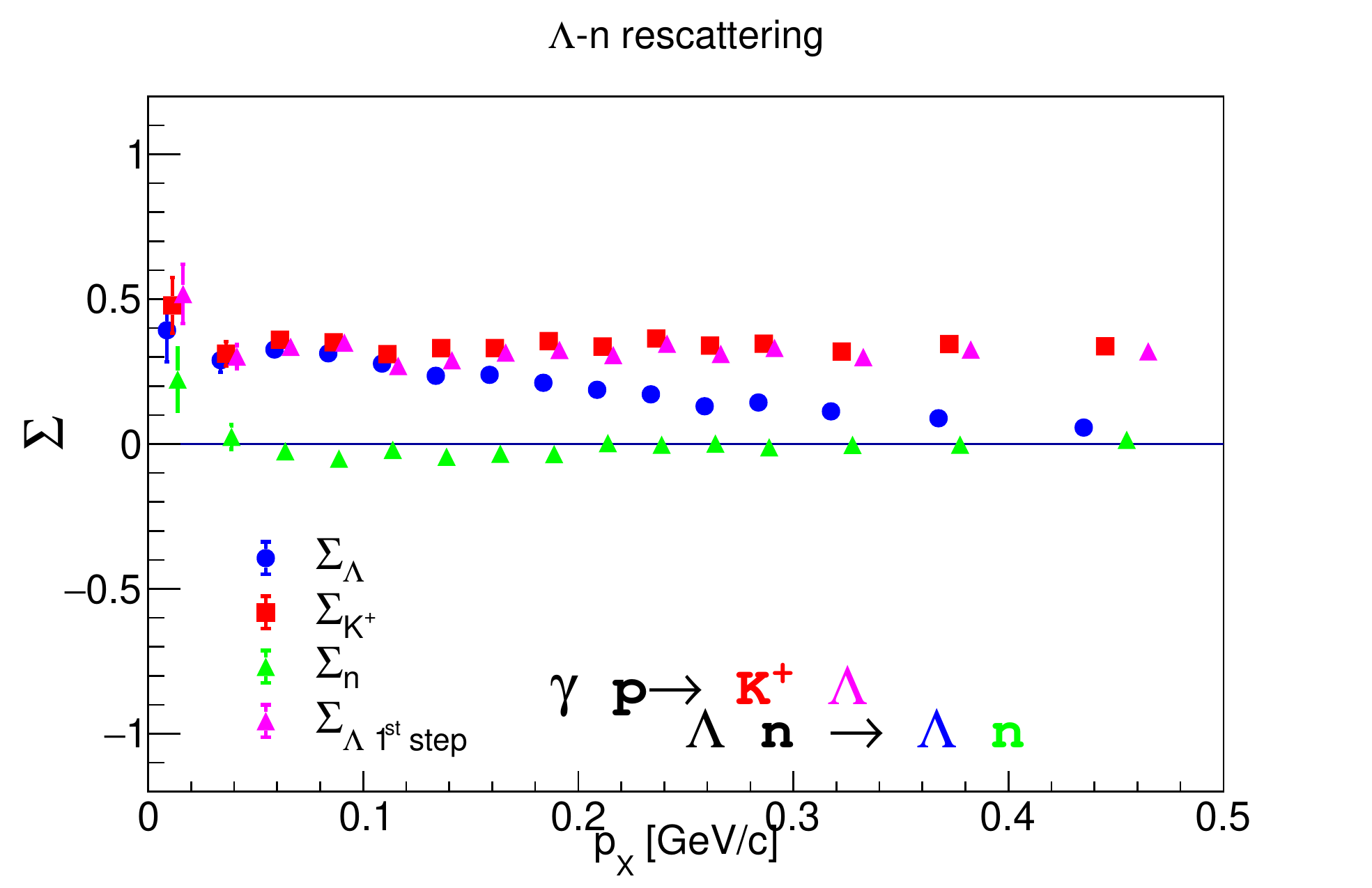}
\caption{\label{Fig:Sigmas} Beam spin asymmetries as a function of the neutron momentum of generated data for the four mechanisms that contribute to the reaction $\gamma d\rightarrow K^+\Lambda n$. The different points show the asymmetries determined using the azimuthal distribution of the various particles as indicated in each caption. }
\end{figure}
Specifically, the kinematic dependence of the beam-spin asymmetry was studied individually for all mechanisms. The beam spin asymmetry of the initial step was set at a specific value and the effect of the second step was investigated. Figure~\ref{Fig:Sigmas} shows the beam spin asymmetries as a function of the neutron momentum for the quasi-free reaction, pion mediated reaction, as well as the Kaon and Hyperon rescattering processes.
 \begin{figure}[h]
\centering
\includegraphics[width=0.57\linewidth]{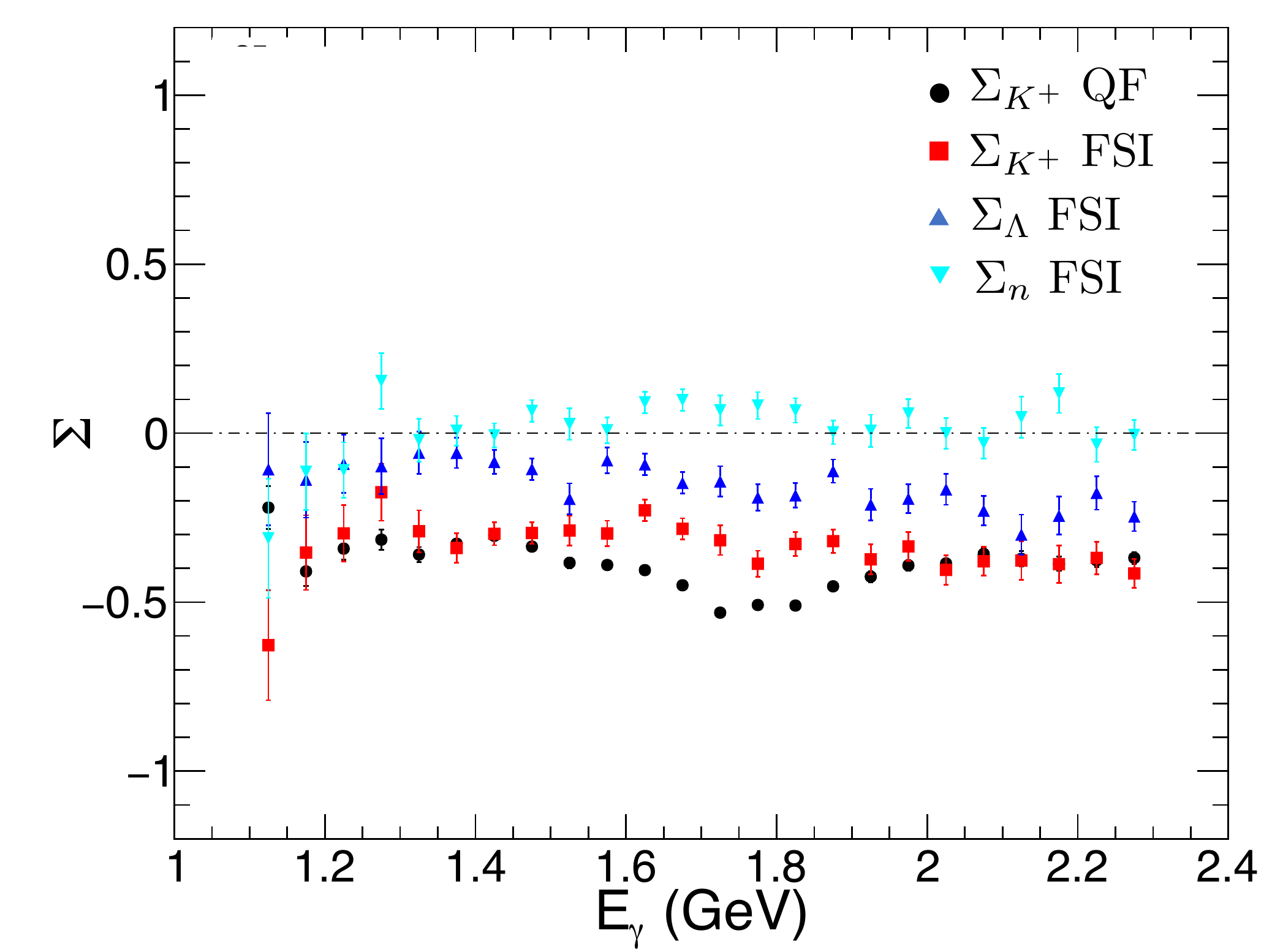}
\caption{\label{Fig:RealSigmas} Beam spin asymmetries as a function of the photon energy of real data. The different points show the asymmetries determined using the azimuthal distribution of the various particles as indicated in the caption. Specifically, the black points show the beam-spin asymmetry of kaons using a QF-dominated sample ($P_x<200~$MeV/c), where as the red blue and cyan points show the beam-spin asymmetries determined using the kaon, hyperon, and neutron azimuthal distribution of an FSI-dominated data sample ($P_x>200~$MeV/c)). }
\end{figure}
 It is evident from these studies that the beam-spin asymmetries determined using the azimuthal distribution of specific particles for each mechanism follows the predicted trend as discussed above. Furthermore, the dilutions are enhanced at missing momenta $P_X$, and a selection of data with missing momenta above 200~MeV/c is expected to to reflect well such dilutions with respect to data with momenta below 200~MeV/c. Detailed studies of simulated data (generated using measurements of the polarised cross section) are well underway to establish the kinematical dependence of the dilutions of the beam-spin asymmetry. This  allow us to obtain the relative ratios of the various FSI mechanisms to the quasi-free production from analysed data from the E06-103 experiment, using determined dilutions from our generated samples. Figure~\ref{Fig:RealSigmas} shows the beam-spin asymmetry of real data using a QF-dominated sample (black points) and FSI dominated samples using the azimuthal distributions of kaon, hyperons and neutrons (red, blue and cyan respectively). 
It is evident that the pion mediated reaction, which results in large $\Sigma_n$ is not a major contributing mechanism in the FSI dominated sample. In addition,  the photon energy regimes  1.1 -- 1.5 GeV and 2 -- 2.3 GeV are dominated by the $YN$ mechanism since $\Sigma_{K^+}$ is consistent with its QF value and $\Sigma_\Lambda$ significantly diluted. Many-fold differential results in these regimes will allows us to place stringent constraints on the underlying dynamics of the YN interaction.

\nolinenumbers

\end{document}